\documentclass[conference]{IEEEtran}
\IEEEoverridecommandlockouts
\usepackage{cite}
\usepackage{amsmath,amssymb,amsfonts}
\usepackage{algorithmic}
\usepackage{graphicx}
\usepackage{textcomp}
\usepackage{hyperref}
\usepackage{xcolor}
\usepackage{url}
\usepackage{balance}

\def\BibTeX{{\rm B\kern-.05em{\sc i\kern-.025em b}\kern-.08em
    T\kern-.1667em\lower.7ex\hbox{E}\kern-.125emX}}
\begin{document}

\title{Rapid rhythmic entrainment in bio-inspired central pattern generators
\thanks{This project has received funding from the European Union’s Horizon 2020 research and innovation programme under the Marie Sklodowska-Curie grant agreement No 101030688, and is partially supported by the Research Council of Norway through its Centres of Excellence scheme, project number 262762.

\copyright 2022 IEEE. Personal use of this material is permitted. Permission from IEEE must be obtained for all other uses, in any current or future media, including reprinting/republishing this material for advertising or promotional purposes, creating new collective works, for resale or redistribution to servers or lists, or reuse of any copyrighted component of this work in other works.}
}

\author{\IEEEauthorblockN{Alex Szorkovszky}
 \IEEEauthorblockA{\textit{RITMO, Department of Informatics} \\
 \textit{University of Oslo}\\
 Oslo, Norway \\
 alexansz@ifi.uio.no}
 \and
 \IEEEauthorblockN{Frank Veenstra}
 \IEEEauthorblockA{\textit{Department of Informatics} \\
 \textit{University of Oslo}\\
 Oslo, Norway \\
 frankvee@ifi.uio.no}
 \and
 \IEEEauthorblockN{Kyrre Glette}
 \IEEEauthorblockA{\textit{RITMO, Department of Informatics} \\
 \textit{University of Oslo}\\
 Oslo, Norway \\
 kyrrehg@ifi.uio.no}
 }

\maketitle

\begin{abstract}

Entrainment of movement to a periodic stimulus is a characteristic intelligent behaviour in humans and an important goal for adaptive robotics. We demonstrate a quadruped central pattern generator (CPG), consisting of modified Matsuoka neurons, that spontaneously adjusts its period of oscillation to that of a periodic input signal. This is done by simple forcing, with the aid of a filtering network as well as a neural model with tonic input-dependent oscillation period. We first use the NSGA3 algorithm to evolve the CPG parameters, using separate fitness functions for period tunability, limb homogeneity and gait stability. Four CPGs, maximizing different weighted averages of the fitness functions, are then selected from the Pareto front and each is used as a basis for optimizing a filter network. Different numbers of neurons are tested for each filter network. We find that period tunability in particular facilitates robust entrainment, that bounding gaits entrain more easily than walking gaits, and that more neurons in the filter network are beneficial for pre-processing input signals. The system that we present can be used in conjunction with sensory feedback to allow low-level adaptive and robust behaviour in walking robots.

\end{abstract}

\begin{IEEEkeywords}
central pattern generator, spiking neuron, entrainment, synchronization, genetic algorithm, robotics, open loop
\end{IEEEkeywords}

\section{Introduction}

Vertebrate locomotion is generally driven by central pattern generators (CPG), distributed networks of locomotor neurons that have evolved to generate oscillatory patterns of movements --- or gaits --- that suit an animal's biomechanics and its environment \cite{ijspeert2008central,guertin2009mammalian}. Typically, signals from the brain stem can modulate the period of these oscillations in order to adjust walking or running speed. A more complex behaviour is the entrainment of movement to an external stimulus, such as playing or dancing to music, or walking in time with a companion \cite{peng2015robotic}.

At least some rhythmic behaviour has been reproduced by continuous-time dynamical systems models of neurons. Two mutually inhibitory neural populations (a ``half-center’’ model) have a period that is easily adjusted by tonic input. Larger numbers of neural populations can trigger the same gait transitions and bistabilities seen in animals \cite{danner2017computational,ausborn2019computational}.

Robots, compared to neurobiological models, generally employ simpler oscillators with fixed frequencies for building CPGs, although adaptive-frequency variations of these have been developed \cite{aoi2017adaptive}. One approach to frequency adaptation is continuous control, in which the frequency is controlled by a time-dependent variable that is continually adjusted according to some error signal. Buchli et al.\ used oscillators with phase errors explicitly fed back into the frequency, so that a robot's gait frequency approached the natural resonance of its passive joints \cite{buchli2006finding}. Iwasaki and Zheng used reciprocal coupling between a half-center CPG and a pendulum to allow synchronization between them \cite{iwasaki2006sensory}. More recently, Egger et al.\ developed a suitable spiking neuron half-center model where the difference between input and output pulses are fed back to the tonic input~\cite{egger2020neural}.

These methods, however, do not fully capture how humans entrain to external rhythms. Evidence from neuroscience suggests that brain oscillations mediate between rhythmic stimulus and the motor system in a top-down fashion, and relax to their normal frequencies after the stimulus ends \cite{calderone2014entrainment,helfrich2019neural}. On the practical side, while the control theory approach can be applied to simple isochronous pulses, it is unclear how to obtain an error function for more complex periodic inputs.

In neuroscience, open-loop models for entrainment have been developed that could in principle be applied to robotics. One is using large recurrent networks, in which time is encoded in the high-dimensional network state \cite{karmarkar2007timing}. Another approach is the gradient frequency network of Large et al., in which an oscillator with matching frequency or harmonic, becomes resonant out of a collection of several \cite{large2010canonical,large2015neural}. It is, however, not clear how to integrate these relatively complex systems with a CPG.

In this paper, we show that a CPG consisting of a small number of non-linear oscillators can rapidly entrain to a complex periodic signal through simple forcing. We demonstrate this idea with a quadruped model based on a network of Matsuoka oscillators that are modified to have input-dependent frequency. The network is composed of a CPG as well an intermediate cortical network that filters the external signal, with unidirectional coupling between the two parts. Parameters are optimized using multi-objective genetic algorithms, generating systems ranging from maximally flexible to maximally stable. We examine some limits of entrainment as a function of signal complexity, signal amplitude and the number of oscillators in the intermediate network.

\begin{figure}
    \centering
    \includegraphics[width=0.46\textwidth]{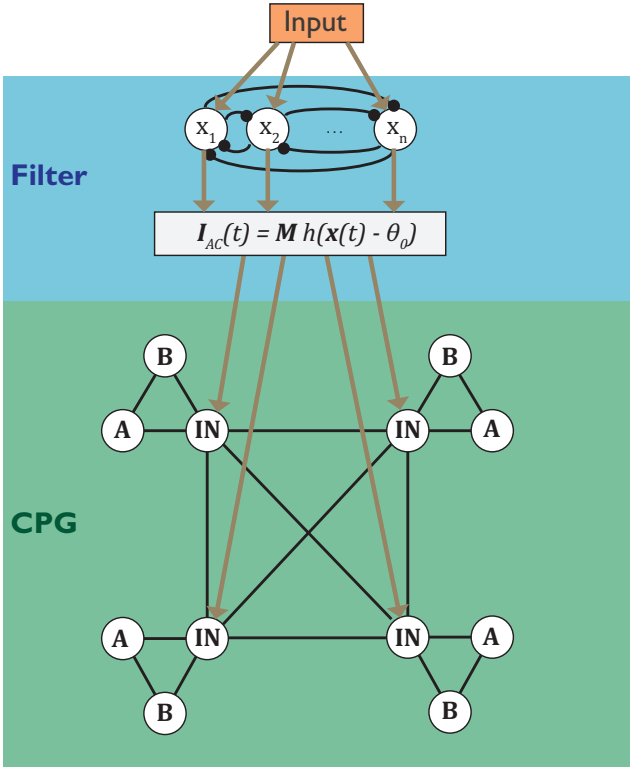}
    \caption{Schematic of the hierarchical neural network. Circles denote modified Matsuoka neurons ($n$ in the filter component, and 12 in the central pattern generator). Arrows denote one-way connections (either excitatory or inhibitory), lines ending in circles denote inhibitory connections, and regular lines denote mutual connections that may be excitatory or inhibitory. A/B: motor neurons; IN: interneurons.}
    \label{fig:diagram}
\end{figure}

\section{Neural Model}

There exists a wide spectrum of mathematical models of oscillators that are commonly applied to central pattern generators. On one end we have the complexity and flexibility of spiking biological models, generally based on variations of the Wilson-Cowan model \cite{ermentrout2010mathematical}. This has recently found great success in closely reproducing the gait transitions of the mouse CPG, including regions of bistability \cite{danner2017computational}. On the other end we have the predictable yet inflexible simple harmonic oscillator, which has been widely used in robotics due to its stability and ease of including feedback. In between there are many levels of variation in non-linearity and degrees of freedom, such as the Hopf oscillator \cite{buchli2006finding}, the Fitzhugh-Nagumo model \cite{shim2018chaotic}, Amari-Hopfield networks \cite{beer1997biologically} and the Rowat-Selverston model \cite{nassour2014multi}.

The Matsuoka neuron is a biologically motivated yet abstract two-variable model \cite{matsuoka1985sustained,taga1991self}:
\begin{align}
t_0 \frac{dx_i}{dt} &= -x_i -ay_i + I_i(t) \\
t_0 \frac{dy_i}{dt} &= -\gamma y_i + bh(x_i)
\end{align}
where $h(x)$ is a rectified linear unit: $h(x)=0$ for $x\leq0$ and $h(x)=x$ for $x> 0$. Like most biological models, there is a fast ``spiking'' variable ($x$) and a slow ``recovery'' variable ($y$). Below a critical value of $a$, a single neuron will not oscillate. However, when two or more neurons are given interconnections using
\begin{equation}
I_i(t) = \sum_j w_{ij}h(x_j - \theta_{ij})
\end{equation}
where $w_{ij}$ and $\theta_{ij}$ are weights and output thresholds, respectively, it is easy to produce oscillatory CPG-like patterns via mutual inhibition or excitation.

The property of a rectified output simplifies the analysis of the phase space. In addition, compared to the Amari-Hopfield model (also sometimes called a continuous-time RNN) \cite{beer1995dynamics}, the region of phase space containing limit cycles is much larger.

However, unlike biological neurons, where the level of constant input can be used to control the firing rate and hence the oscillation frequency, the firing rate of the Matsuoka model is insensitive to tonic input \cite{jouaiti2019comparative}. Hence, we introduce a sigmoidal activation function $S(x)$ akin to those used to model persistent sodium currents in motor neurons:
\begin{align}
t_0 \frac{dx_i}{dt} &= -x_i -aS(\kappa[x_i-x_0]) y_i + c_i + d_i I_{DC} + I_{ACi}(t) \\
t_0 \frac{dy_i}{dt} &= -\gamma y_i + bh(x_i)
\end{align}
where $S(x)=1/(1+\exp(x))$. Here $d_i$ is the coefficient for the tonic brain stem drive $I_{DC}$ used to tune the gait period. When the coefficients satisfy:
\begin{equation}
c_i + d_i I_{DC} > x_0 + \frac{2}{k}   \; , 
\end{equation}
this reproduces the ubiquitous cubic-like shape of the fast variable's nullcline \cite{skinner1994mechanisms}. In this case, each neuron may self-oscillate for a certain parameter range. 


\section{CPG Model}

We assembled a quadruped CPG as a modular system of limb controllers connected by interneurons, in a similar fashion to Beer\cite{beer1997biologically} and Ijspeert\cite{ijspeert2001connectionist}. 
For an overview, see Fig.~\ref{fig:diagram}.
Each limb controller contains a single interneuron and two motor neurons (`A' and `B'), the latter of which can be used for a single extensor-flexor pair, or for a pair of single-variable joints. Each of the three neurons has its own bias $c_i$ and drive coefficient $d_i$, which are identical for all modules. Thresholds between all CPG neurons $\theta_{ij}$ are zero, and connection weights $w_{ij}$ are zero between modules, apart from when $i$ and $j$ are both interneurons. The CPG has lateral symmetry, so that connection weights are equal for equivalent connections between the left and right side of the body.

\begin{table}
\centering
\caption{Parameter ranges for the CPG network.}
\label{paramtable}
\begin{tabular}{c|c}
Parameter & Value / Range \\
\hline
$t_0$       &  0.01 s\\
$\gamma$    & [0.01,0.1] \\
$a$         & [0.2,2]    \\
$b$         & [0.02,0.2] \\
$\kappa$    & [0.5,5] \\
$x_0$       & [0.1,1] \\
$d_i$       & [-0.9,0.9] \\
$c_i$       & [1.1,2] \\
$w_{ij}$    & [-1.8,1.8] / 0\\
\end{tabular}
\end{table}

\subsection{CPG Optimization}

The CPG was implemented in Python\footnote{Genotypes of individuals in this paper and source code to generate all results are available at 
\url{https://github.com/aszorko/COROBOREES/tree/Paper1}.} 
and optimized using the NSGA3 genetic algorithm \cite{deb2013evolutionary} included in the DEAP toolbox \cite{DEAP_JMLR2012}. Each parameter that is not set according to the above constraints is encoded by an integer between 1 and 10, which determines its value within the range shown in Table \ref{paramtable}. Note that $c_i$ and $w_{ij}$ (for connections that exist according to the schematic) cannot be zero. In addition, connections between interneurons are constrained to be inhibitory ($w_{ij}<0$).

A given CPG was evaluated by iterating the brainstem drive $I_{DC}$ from 0 to 1 in steps of $0.1$. For each drive (indexed by $k$), the system was given random initial conditions, followed by a burn-in period, after which the time series of a flexor-extensor-type output (difference between rectified A and B neuron outputs) was analysed for each of the four limb modules. First, the periods were measured using the maximum of the autocorrelation function, and reduced to the mean oscillation period $T_k$ and coefficient of variation $\mathrm{CV}_{Tk}$ over the four limbs. A measured correlation peak at a time less than $0.01$s was considered non-oscillating and hence invalid. A period shift of $|T_{k+1}-T_k|/(T_{k+1}+T_k)>0.15$ was also considered invalid in order to filter out large discontinuities. In addition, the mean oscillation amplitude $A_k$ and coefficient of variation $\mathrm{CV}_{Ak}$ were measured using a peak finding algorithm on these same time series. Mean amplitudes less than $0.1$ or greater than 10 were also considered invalid in order to keep all CPG outputs within a comparable range. Finally, ``duty functions'' $D_{Ak}$ and $D_{Bk}$ were measured to penalise unbalanced gaits in which three or more limbs are activated simultaneously:
\begin{equation}
D_{A} = E_t\left[\sum_i^4 I\left(\left|\frac{d}{dt} h(x_{Ai})\right|>\epsilon\right) < 3\right] \; ,
\end{equation}
and similarly for $D_B$, where $E_t$ is the time domain expectation value, $I$ is the binary indicator function, and $\epsilon=0.001$ in this study. If there is no oscillation for a given $k$ then both $D_{Ak}$ and $D_{Bk}$ are given a value of zero. The use of the derivative was to allow the possibility of consistent flat output, while penalizing simultaneous spiking of equivalent neurons in three or more limbs.

Three fitness functions were to be maximized by the multi-objective optimization:
\begin{align}
F_1 &= \left|\sum_{k=1}^{N-1} V_k V_{k+1}\frac{T_{k+1}-T_k}{T_\mathrm{max}}\right| \\
F_2 &= \frac{1}{N}\sum_k^N \frac{1}{1 + \mathrm{CV}_{Tk} + \mathrm{CV}_{Ak}} \\
F_3 &= \frac{1}{2N}\sum_k^N D_{Ak} + D_{Bk}\\
\end{align}
where $V_k$ is the validity of the time series during drive $k$, averaged over the four outputs, with each validity being 0~or~1. 

NSGA3 generates a Pareto front containing non-dominated individuals according to these fitnesses, thus selecting for CPGs with some combination of large monotonic variation in period as a function of $I_{DC}$, low variation between limbs, and two or more limbs consistently in the ``stance'' segment of the cycle.

We used a population of 48 individuals that evolved using two-point crossover and mutation for 100 generations. Each individual was evaluated again 5 times in order to use medians as accurate final fitnesses, and from these a final Pareto front of CPGs was generated. Four solutions were selected in a way that balanced variety and average overall performance. First the highest overall fitness was selected using the sum of fitnesses. Then, three more individuals were selected using the maxima of
\begin{equation}
F_m^* = zF_m + \sum_{k=1}^3 F_k
\end{equation}
where $z$ was incremented in intervals of one until the maximum of each $F_m^*$ was unique.

\begin{table}
\centering
\caption{Parameter ranges for the filter network.}
\label{filtparamtable}
\begin{tabular}{c|c}
Parameter & Value / Range \\
\hline
$t_0$       &   0.01 s\\
$\gamma$    &   0.03 \\
$a$         &   2    \\
$b$         &   0.3 \\
$\kappa$    &   4 \\
$x_0$       &   1 \\
$d_i$       &   0 \\
$\theta_0$    &   0.15 \\
$\Gamma$    &   [0.05,0.55] \\
$c_i$       &   [2,2.5] \\
$G_{i}$    &   [-1,1] \\
$w_{ij}$    &   [-6/(n-1),0] \\
$M_{ij}$    &   [-10,10] \\
\end{tabular}
\end{table}

\begin{table*}[h]
\centering
\caption{Top CPGs after 100 generations of selection, and mean fitnesses $F_f$ of each CPG's best filter. All values $F_i$ and $F_{fi}$ are the medians from five evaluations.}
\label{cpgresults}
\begin{tabular}{c|c|c|c|c|c|c|c|c}
CPG & Optimized & $T_{0.5}$(s) & $F_1$ & $F_2$ & $F_3$ & Gait & $F_f(n=2)$ & $F_f(n=4)$\\
\hline
0 & Overall & 1.80 & 0.76 & 1.00 & 0.68 & bound  & 0.96 & 0.97 \\
1 & $F_1^*$   & 1.35 & 0.76 & 0.70 & 0.79 & walk & 0.33 & 0.53 \\
2 & $F_2^*$   & 0.47 & 0.04 & 0.98 & 1.00 & walk & 0.32 & 0.30 \\
3 & $F_3^*$   & 1.15 & 0.63 & 0.68 & 0.91 & walk & 0.32 & 0.95  \\

\end{tabular}
\end{table*}

\section{Filter Network}

For each of the three selected CPGs, a filter was evolved on top. The purpose of the filter is to pre-process the input and distribute the signal among the CPG modules as neuron-like pulses. The filter network was a single non-lateralized layer of $n$ neurons. The coupling to the input was then governed by $n$ coefficients. In order to not disturb the CPG in the absence of input, the neuron parameters were set to be below the spontaneous bursting threshold, and all interconnections were made to be inhibitory ($w_{ij} \leq 0$). However, this does not entirely preclude oscillation of the network, and so further measures were taken in the fitness function below. From the input, $n$ coefficients $G_i$ govern the coupling to the neurons, while 4$n$ coefficients governed the coupling from the filter to the four CPG interneurons (forming the matrix M in Fig.~\ref{fig:diagram}). For these connections, the offset $\theta_{ij}$ was set to a constant value $\theta_0$ to offset the equilibrium output of the neurons.

\subsection{Filter Network Optimization}

The input consisted of spikes at a regular time interval with period $\tau$, which was then low-pass filtered using an exponentially decaying impulse response with decay constant $\Gamma / t_0$. The periods $\tau_k$ used were $2/3$, $1$ and $3/2$ multiplied by $T_{0.5}$, where $T_{0.5}$ is the median period for a tonic input of $I_{DC} = 0.5$. Together these present a range of more than a factor of two in input period. Each neuron in the filter network received the input multiplied by its own coefficient $F_i$.

The network was evolved to minimise the mean difference between the input period and the periods of the motor neurons. Although this is a relatively simple objective function, NSGA3 was used again to avoid early convergence to a local optimum. The period $T_{ik}$ of the four CPG motor neurons was measured as in the previous section, and the NSGA3 algorithm was run using three fitness functions, one for each $\tau_k$:
\begin{equation}
F_{fk} = \frac{V_k}{1 + \sigma_0 / \sigma_t + \sqrt{\sum_i{(T_{ik}-\tau_k)^2}/4}}
\end{equation}
where the validity $V_k$ is defined as in the previous section, $\sigma_0$ is the mean standard deviation of the filter output with no input and $\sigma_t$ is a scaling threshold, set to $0.1$ for the current study. A population of 68 was evolved for 50 generations for filters with $n=4$ neurons, and for 25 generations for $n=2$ due to the much smaller number of parameters. The filter with the highest overall sum of fitnesses was then chosen for analysis.

Finally, for the overall best CPG, an additional filter was evolved to test the ability to entrain to more complex signals \cite{tal2017neural}. This was done in the same way as the original filter, but with every fourth pulse missing from the input.

\section{Results}

\begin{figure}
    \centering
    \includegraphics[width=0.45\textwidth]{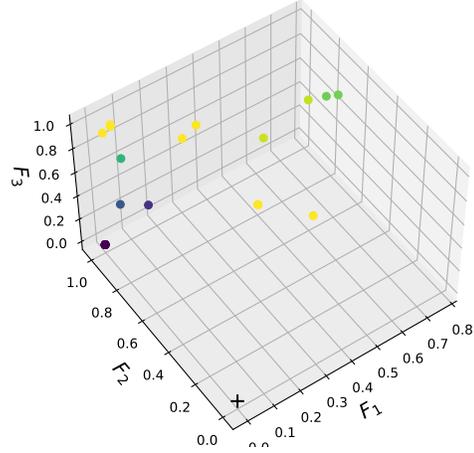}
    \caption{Median fitnesses of the individual CPGs in the Pareto front after 100 generations. Lighter color indicates higher vertical position ($F_3$).}
    \label{fig:cpgevo}
\end{figure}

\begin{figure}
    \centering
    \includegraphics[width=0.45\textwidth]{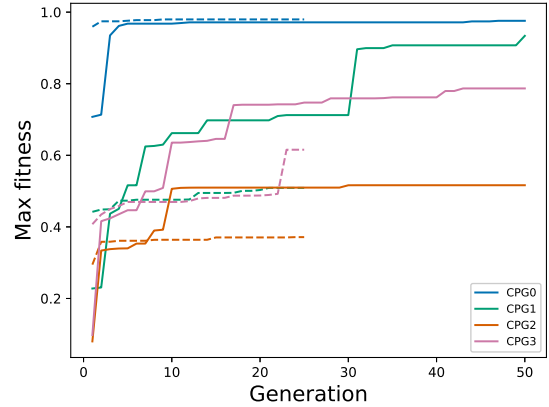}
    \caption{Maximum fitnesses of the population during evolution of the filter modules, averaged over the three $F_{fk}$ values. Solid lines: $n=4$, dotted lines: $n=2$.}
    \label{fig:filtevo}
\end{figure}

The three CPG fitnesses reached stable levels within 100 generations, with the final Pareto front shown in Fig.\ \ref{fig:cpgevo}. Results for CPGs maximizing the four weighted averages are shown in Table \ref{cpgresults}. The overall highest mean fitness was achieved by a bounding gait, with front and hind limb pairs moving in synchrony, while a variety of walking gaits maximized the fitnesses weighted towards individual components. The four CPGs encompassed a wide range of periods, from $0.47$ to $1.8$ seconds.

\begin{figure}
    \centering
    \includegraphics[width=0.45\textwidth]{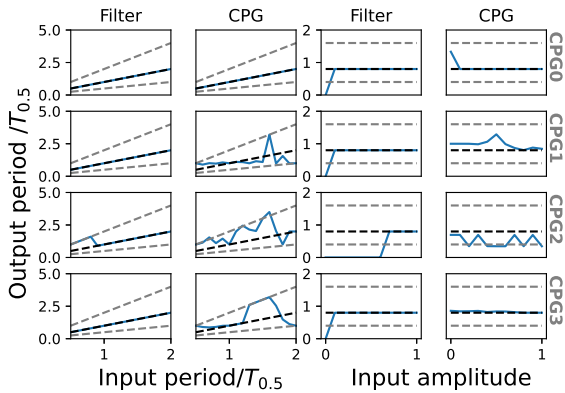}
    \caption{Median periods of the filter and CPG outputs relative to the CPG frequency $T_{0.5}$, as a function of input period (amplitude $=1$) and amplitude (period $=0.8\times T_{0.5}$) for $n=4$. Black dotted lines represent the input period, while grey dotted lines represent half and double the input period. All values are a single evaluation.}
    \label{fig:periods}
\end{figure}

The CPG with highest overall fitness was also the most successful at entrainment, with fitness close to the maximum of one for both 2-neuron and 4-neuron filter configurations, as shown in Fig.\ \ref{fig:filtevo}.  The full CPG0 system was able to generalize beyond the input periods and below the amplitude used for training, as shown in Figure \ref{fig:periods}. Notably, the CPG with the least flexible period was least able to evolve a filter to entrain the CPG at other drive periods. The walking gaits in general had more difficulty generalizing entrainment to arbitrary periods, sometimes entraining at a multiple of the input period but often having periods not associated with the input, suggesting highly nonlinear behaviour. Period doubling was also seen in the filter of CPG2, which is an expected outcome of driving the systems at high frequencies beyond those used for training.

\begin{figure}
    \centering
    \includegraphics[width=0.45\textwidth]{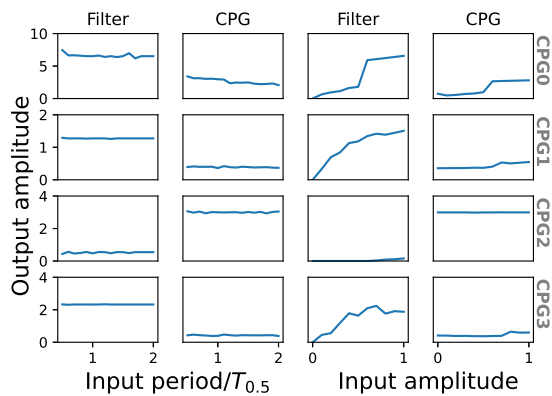}
    \caption{Mean peak-to-peak output amplitudes of the filter and CPG as a function of input period (amplitude$=1$) and amplitude (period=$0.8\times T_{0.5}$) for $n=4$. All values are a single evaluation}
    \label{fig:amps}
\end{figure}

The amplitude profiles as a function of input period were also mostly flat as shown in Figure \ref{fig:amps}, implying very wide transfer functions. The output amplitude curves, however, show substantial jumps in three of the CPGs when the input amplitude reaches a certain threshold. This can be driven either by the filter or the CPG dynamics.


\begin{figure*}
    \centering
    \includegraphics[width=0.7\textwidth]{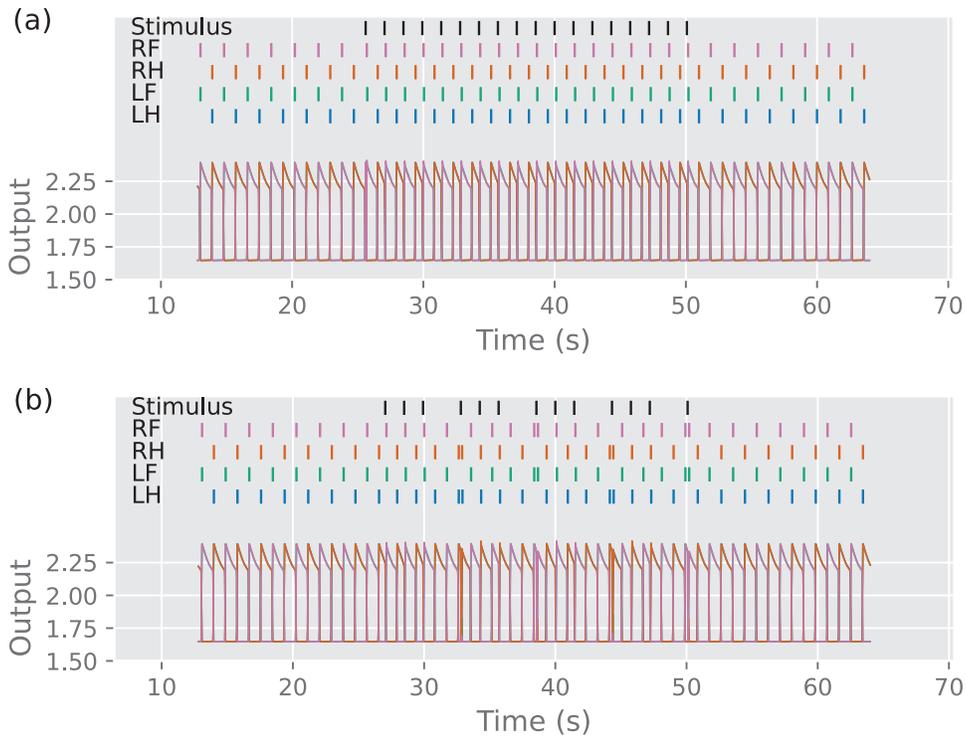}
  \caption{CPG0 combined output ($h(x_{Bi})-h(x_{Ai})$) for (a) a transient isochronous input with $\tau=0.8\times T_{0.5}$, amplitude$=0.7$ and (b) the same input with every fourth pulse missing, both using the filter trained on the missing pulse. Time series for the LF (left front) and LH (left hind) limbs lie behind the RF (right front) and RH (right hind) limbs, respectively. Ticks above show locations of peaks.}
    \label{fig:timeseries}
\end{figure*}

Filters were then evolved for CPG0 using a non-isochronous input. The resulting fitness (median $F_f=0.61$ for $n=4$, $F_f=0.39$ for $n=2$) were lower than for the filter evolved for isochronous input. However, these filter performs reasonably well on isochronous input (median $F_f=0.64$ for $n=4$, $F_f=0.91$ for $n=2$), and better than the original CPG0 filters on non-isochronous input (median $F_f=0.30$ for $n=4$, $F_f=0.33$ for $n=2$). As shown in Figure \ref{fig:timeseries}, in both cases the $n=4$ system adjusts its period rapidly to the stimulus, retaining its original gait pattern, and relaxing back to its original period after the stimulus ends.

\section{Discussion}

To our knowledge, this is the first demonstration of open-loop rhythmic entrainment in a central pattern generator model. This was achieved via a novel modification of a well-used neural model, with evolutionary optimization not only of connection weights but also of several system parameters that determine the degree of nonlinearity. In this architecture, wide tunability of the oscillation period with tonic input appears to be a precondition for synchronization to a wide range of periodic inputs. The neural model we develop captures this important property while remaining simple enough for use in robotic applications.

Our results demonstrate that only a small total number of neurons (16) are required for robust entrainment compared to the number of neuron populations in biological quadruped CPGs (with recent modelling studies involving several dozen \cite{danner2017computational}). However, the difference in results between two-neuron and four-neuron filters suggests that more neurons will add robustness and flexibility for more complex gaits (such as quadruped walking), transitioning gaits, and more complex inputs.

Although communication between the brain and body is clearly two-way in animals, this work presents a complementary approach to closed-loop control methods for temporal prediction. The synchronization-based approach circumvents the problem of determining a suitable error function for complex temporal patterns. For entrainment, feedback may take on a fine-tuning role, as is hinted at by brain research that differentiates processes occurring on long and short time-scales \cite{karmarkar2007timing}.

Our conceptual model has applications both within neurophysiological research and in the design of intelligent systems. For the former, our framework can be further developed to investigate general principles behind rhythmic entrainment and embodied cognition \cite{helfrich2019neural,wilson2013embodied}. Practical applications include beat tracking \cite{large1995beat} and adaptive and social robotics ~\cite{peng2015robotic,mortl2014rhythm,nocentini2019survey,khoramshahi2019dynamical}.

In the future, this system will be tested in simulated and physical robots. The addition of sensory and balancing feedback is expected to improve the overall stability of the system. The real-time low-level adaptive behaviour that we demonstrate also opens up the possibility of realistic human-robot and robot-robot interaction. To this end, studies with multiple, mutually interacting agents will allow the study of emerging collective behaviors.

\bibliographystyle{unsrt}
\balance
\bibliography{main.bib}

\end{document}